# ACCELERATOR COMPLEX EVOLUTION AT FERMILAB

M. Convery†, Fermi National Accelerator Laboratory, Batavia, United States

*Abstract*

The largest hadron accelerator facility in the US is undergoing radical changes and the undertaking of new HEP-driven neutrino research. This talk will discuss the wide-ranging projects and impacts to the accelerator community taking place at FNAL.

## INTRODUCTION

For roughly the last decade, the Fermilab accelerator complex has been supporting world-class short- and long-baseline neutrino, muon, and test-beam programs.

As Fermilab's main focus shifts to the Deep Underground Neutrino Experiment (DUNE), the accelerator complex must modernize to improve reliability and maximize beam power. The 1970's Linac will be replaced by a new superconducting-RF (SRF) linac, Linac2, already under construction. The Booster will be upgraded to cycle at 20 Hz and to optimize for beam from the new linac. The Main Injector must reliably provide 1.2-2.4 MW proton beams to the new Long Baseline Neutrino Facility (LBNF) to produce neutrinos for DUNE.

## STATUS OF THE FERMILAB COMPLEX

The Booster Neutrino Beam has delivered $3\times10^{21}$ protons on target (POT) at 8 GeV to the short-baseline neutrino experiments over the last ten years including the requested $6.6\times10^{20}$ POT to the Short Baseline Near Detector, yielding the largest neutrino-argon dataset in the world. Two additional runs are planned prior to the Linac2 era.

The Main Injector, which delivers 120-GeV proton beam for the long-baseline neutrino experiments, has been executing a campaign to increase beam power by shortening the cycle time rather than by increasing beam intensity. The ramp was shortened from an initial 1.33s to 1.067s, and a record beam power of 1.02 MW was demonstrated in 2024 [1].

The Muon Campus has been commissioning resonant extraction for the Mu2e experiment in preparation for a physics run which is planned for late 2027.

Infrastructure failures in 2024, the most notable of which was an electrical transformer needed to power the Main Injector, brought to the forefront the need to invest in reliable infrastructure for the DUNE era. In spite of recent challenges, we have delivered $4.6\times10^{21}$ POT to the NOvA experiment since 2014. The Accelerator and NuMI Upgrades portion of the NOvA project was designed to deliver $3.6\times10^{21}$ POT *(over a 6-year period).*

†convery@fnal.gov

## TRANSITION TO THE LINAC2, LBNF/DUNE ERA

A comparison of the Accelerator Complex in 2026 and in 2031 is shown in Figs. 1-2.

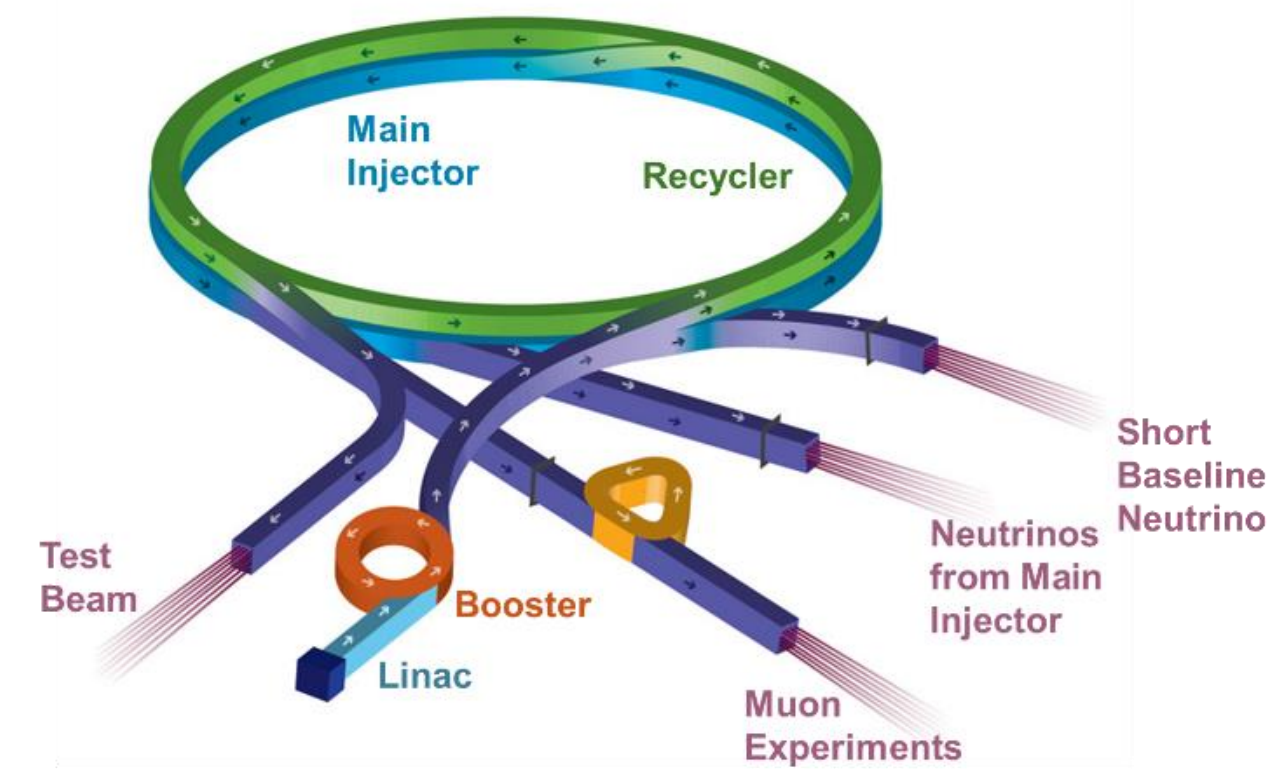

Figure 1: Fermilab Accelerator Complex in 2026

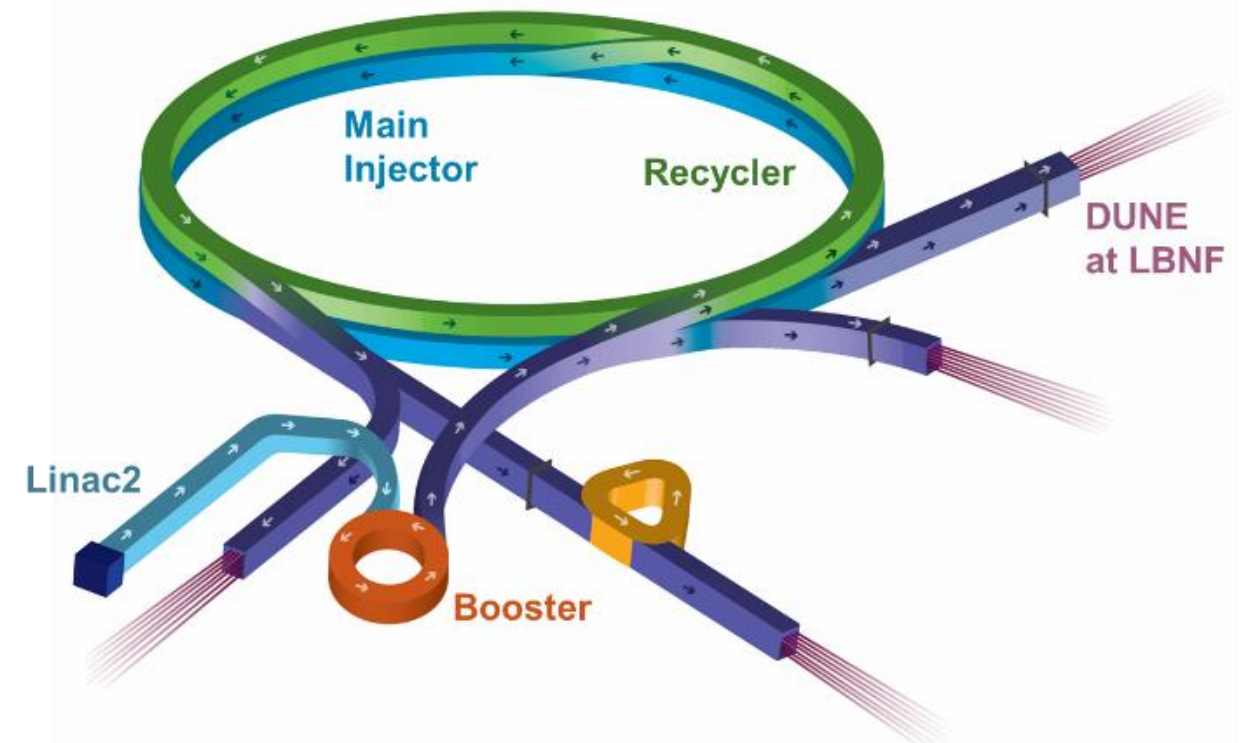

Figure 2: Fermilab Accelerator Complex in 2031.

The Proton Improvement Plan II (PIP-II) project is building a state-of-the-art superconducting-RF linac, Linac2, that will accelerate H- ions to 800 MeV [2]. A new Beam Transfer Line (BTL) will connect to the existing Booster, which will be upgraded to deliver the 8-GeV proton beam at a rate of 20 Hz, compared to the current 15 Hz. Together with other upgrades of the Booster and Main Injector, this will provide a beam power of 1.2 MW to LBNF/DUNE.

The LBNF/DUNE-US project provides a beamline to deliver a 1.2-2.4 MW proton beam, a target station to produce neutrinos, a "near detector" facility on the Fermilab site in Illinois, and excavated and outfitted detector caverns 1.5 km underground at the Sanford Underground Research Facility in South Dakota for the "far detector". The international DUNE collaboration provides the detectors with support from the US project.

The PIP-II project began construction of the new SRF Linac2 in 2023. In 2028, the user program will turn off for two years to connect the Linac2 to the Booster. Construction to connect the new LBNF beamline to the Main Injector is also planned to occur during this time.

## PIP-II INTEGRATION

Some accelerator systems scope is needed for operation of the complex with the Linac2 beyond what is provided by the PIP-II project [3].

Existing readout for instrumentation such as beam position monitors and beam loss monitors throughout the complex uses obsolete electronics and must be upgraded for 20-Hz operation. A prototype has been completed and is being tested with beam.

The injection time for 2-mA beam from the Linac2, compared to ~20 mA from the Linac, increases to ~550 µs. The Booster operates as a resonant circuit synchrotron, where the bending field ramps sinusoidally at 15 Hz, to be upgraded to 20 Hz, and the magnetic field changes significantly during the longer injection time. Using corrector magnets, a 400-µs long flat region has been implemented so far, and power supply upgrades for the corrector magnets are planned [4].

Of concern was the possibility that operating the Booster gradient magnets at higher voltage needed for 20 Hz could result in corona discharge, which could cause them to fail over time. Partial discharge measurements have shown that some magnets have been experiencing some level of corona discharge already during 15-Hz operation for an extended period of time. Simulations indicate that the voltage to ground expected when running at 20 Hz will not dramatically increase the severity of the discharges or the frequency at which they occur. Observations have also shown that magnets which have been subjected to high integrated radiation fields do not necessarily present more severe corona discharges or lower thresholds than those which have never been in the beam line. Widespread magnet failures due to electrical discharges are therefore unlikely within the first few years of 20-Hz operations. Spare magnets could be fabricated to mitigate the risk.

The PIP-II project will build and install five larger-bore RF cavities in place of existing Booster cavities. Together with the one already installed by the original Proton Improvement Plan, this will provide sufficient RF voltage to meet the project requirements. An additional six are needed for full beam operations plus six more for reliability. The plan is to continue the construction of new cavities after the ones for PIP-II are complete.

In the Linac2 era, with the existing cavity tuners, the injection bias current would have to be increased. For Booster operations at 20 Hz and with the increased injection bias current, the existing busbars would heat up beyond what their environment can support. The most cost-effective solution is to replace the cavity tuners with new tuners with a lower required bias current.

As presently designed, the Beam Transport Line (BTL) connecting Linac2 to Booster does not have any capability to rotate the longitudinal phase space of the beam to match the acceptance of the Booster. At 1.2-MW operation, when the intensity will be high and space-charge effects will be important, it is necessary to have a buncher cavity in the BTL to match into the Booster. This buncher can also compensate for loss of a cavity in the half-wave resonator of the Linac2 to keep the momentum spread of the beam within acceptable limits.

Also needed for 1.2-MW operation with a Booster intensity of $6.7 \times 10^{12}$, is a $\gamma_t$-jump system to reduce beam losses during transition crossing in the Booster. Scope could include new power supplies for quadrupole magnets and Inconel beampipe. Note that the PIP-II project provides a $\gamma_t$-jump system in the Main Injector [5].

## ACCELERATOR INFRASTRUCTURE READINESS

Infrastructure failures in 2024 brought to the forefront the need to invest in reliable infrastructure for the DUNE era. Although Fermilab recognized the need to invest in aging water and electrical systems, limited available funding was more often directed towards operating existing systems.

The Fermilab electrical system is owned and operated by the laboratory. Two external 345-kV transmission lines feed two electrical substations. Each substation contains four high voltage power transformers that step down the incoming power to 13.8 kV for local distribution throughout the site. A 13.8-kV transformer caught fire in 2024 and must be replaced. In the course of investigating, three other transformers had bushings which were found to require replacement. Four of the eight transformers are more than 50 years old, and all are more than 20 years old. A replacement program is being devised. Lower-level transformers and voltage switches for the Booster and Main Injector are also end-of-life and are planned for replacement.

A significant degradation of the units where pond water is used to exchange heat with the low-conductivity water (LCW) that cools magnets and power supplies for the Main Injector was also observed in 2024. The heat exchangers have since been repaired or replaced. End-of-life sump pumps in the Booster, which have been a significant source of downtime, have been replaced, and degraded sump discharge lines for the Booster and Main Injector are also being replaced. Addressing aging pond pumps, air conditioning for equipment cooling, and tunnel dehumidification is also planned.

Operation of the Booster at 20 Hz in the Linac2 era will generate more heat and require an upgrade of the air conditioning in the Booster gallery as well as the LCW system to provide increased flow to the RF. The challenge is finding space for added equipment in the already crowded gallery. The LCW work must be completed during the planned shutdown for connecting the Linac2 to the Booster.

## MAIN INJECTOR RAMP AND TARGETRY

The plan for LBNF/DUNE is to initially take beam at 1.2 MW and later upgrade to 2-2.4 MW. Although a doubling of beam power was originally envisioned via a replacement accelerator for the Booster which could double the beam intensity, the idea of instead increasing the Main Injector ramp rate has been explored as a less expensive means for achieving higher beam power. Indeed, a record beam power from the Main Injector of 1.02 MW was achieved by reducing the cycle time to 1.067s.

While beam is being accelerated in the Main Injector and extracted, 12 batches of protons from the Booster are being slip-stacked together in the Recycler for the next cycle [6]. With beam from the Booster at 20 Hz, the time needed for this is 0.65s. The shortest possible cycle time is therefore achieved by ramping the Main Injector in 0.65s instead of the nominal 1.2s, which increases the beam power from 1.2 MW to greater than 2 MW. The MIRT program would increase the ramp rate by modifying the magnet power supplies and RF systems.

Replacing magnet power supplies with a new, more efficient, design would allow added voltage without expanding or adding new service buildings. Replacing a subset of RF cavities with a new second-harmonic design shows promise, and additional studies are in progress. Tuneable superconducting RF is also being explored. The cost/benefit may be significantly better for an intermediate cycle time such as 0.9s where additional RF cavities of the existing design can fit in the available space.

The LBNF/DUNE-US project provides a target complex with shielding and other infrastructure that can support at 2.4-MW beam, but with target and focusing horns designed to withstand a 1.2-MW beam [7]. The MIRT program will also upgrade the target systems for up to 2.4 MW. The horns will be analysed to see how much beam power they can withstand, and new horns will be designed as needed. For targets, an R&D program is being conducted to find materials suitable for higher beam power [8].

## ACCELERATOR CONTROLS

The Accelerator Complex Operations Research Network (ACORN) project will modernize the Fermilab accelerator controls system [9]. The new system will be allow using EPICS and leveraging experience from other US national labs. PIP-II has already introduced the use of EPICS. A staged plan will address components in order of priority based on requirement for operating with Linac2 (timing system), risk, need of a long shutdown, end-of-life or obsolescence, and impact. The existing ACNET controls system which was developed in the 1980's will continue to be used in parallel until all areas are upgraded.

Data acquisition and control systems will be upgraded to replace obsolete rack monitors and CAMAC cards. A new timing system will be provided for operation at 20 Hz and synchronization with the Linac2. Computing services infrastructure will be upgraded to provide a secure foundation for integrating the existing ACNET and new EPICS platforms, and will facilitate deployment of AI tools. Modern user applications will also be written and deployed.

The ACORN project has also looked at replacing power supplies which have reached end of life and voltage regulation systems in collaboration with Oak Ridge National Laboratory. New LBNF components should use modern technology, so design of a voltage regulation system for magnet power supplies and controls for corrector magnet power supplies are a priority. Ferrite bias supplies for RF in the Booster and Main Injector are obsolete and have limited spares. A replacement will be designed and prototyped by the ACORN project for replacement of supplies as needed and as funding allows.

Some controls work is best done during the long shutdown for tying in Linac2 and LBNF. An example is the ramp control hardware and software for the Main Injector. Others can be phased in gradually or during shorter shutdowns. Critical systems should be upgraded in time for commissioning with Linac2 and LBNF.

## CONCLUSION

The Fermilab accelerator complex is undergoing an evolution to the LBNF/DUNE era, fuelled by the new SRF Linac2. Updated infrastructure and a modern controls system will improve the reliability and efficiency of the complex. New systems will be put into place to support 20-Hz operations with the Linac2. Looking forward to the next phase of increasing beam power to DUNE from 1.2 MW to 2-2.4 MW, increasing the Main Injector ramp rate instead of increasing beam intensity seems to be a promising alternative that would not require constructing a new Booster.

## ACKNOWLEDGMENTS

This manuscript has been authored by Fermi Forward Discovery Group, LLC under Contract No. 89243024CSC000002 with the U.S. Department of Energy, Office of Science, Office of High Energy Physics.